\documentclass[
superscriptaddress,
twocolumn,
longbibliography,
pra,
aps,
10pt
]{revtex4-1}
\usepackage{graphicx}
\usepackage{amsmath,amsfonts,amssymb}
\usepackage{xcolor}
\usepackage{hyperref}
\hypersetup{colorlinks=true,allcolors=blue}
\usepackage[T1]{fontenc}
\usepackage{footnote}
\usepackage{siunitx}
\usepackage{physics}
\usepackage{dsfont}
\usepackage{tabularx,array}

\usepackage{xcolor}
\definecolor{color_g}{HTML}{2577B0}
\definecolor{color_dm}{HTML}{277A28}
\definecolor{color_bm}{HTML}{FD822B}
\definecolor{color_b}{HTML}{B4B4B4}
\definecolor{color_bp}{HTML}{B4B4B4}
\definecolor{color_dp}{HTML}{B4B4B4}
\definecolor{color_text}{HTML}{B4B4B4}

\renewcommand{\vec}[1]{\mathbf{#1}}

\begin{document}

\title{Phonon wave packet emission during state preparation of a semiconductor quantum dot using different schemes}
\author{Thomas K. Bracht*}
\affiliation{Institut f{\"u}r Festk{\"o}rpertheorie, Universit{\"a}t M{\"u}nster, 48149 M{\"u}nster, Germany}
\author{Tim Seidelmann}
\affiliation{Theoretische Physik III, Universit{\"a}t Bayreuth, 95440 Bayreuth, Germany}
\author{Tilmann Kuhn}
\affiliation{Institut f{\"u}r Festk{\"o}rpertheorie, Universit{\"a}t M{\"u}nster, 48149 M{\"u}nster, Germany}
\author{V. Martin Axt}
\affiliation{Theoretische Physik III, Universit{\"a}t Bayreuth, 95440 Bayreuth, Germany}
\author{Doris E. Reiter}
\affiliation{Institut f{\"u}r Festk{\"o}rpertheorie, Universit{\"a}t M{\"u}nster, 48149 M{\"u}nster, Germany}

\date{\today}

\begin{abstract}
The carrier-phonon interaction in semiconductor quantum dots can greatly affect the optical preparation of the excited state. For resonant excitation used in the Rabi preparation scheme, the polaron is formed accompanied by the emission of a phonon wave packet, leading to a degradation of preparation fidelity. In this paper, phonon wave packets for different coherent excitation schemes are analyzed. One example is the adiabatic rapid passage scheme relying on a chirped excitation. Here, also a phonon wave packet is emitted, but the preparation fidelity can still be approximately unity. A focus is on the phonon impact on a recently proposed swing-up scheme, induced by two detuned pulses. Similar to the Rabi scheme, a degradation and a phonon wave packet emission is found, despite the detuning. If the swing-up frequency coincides with the maximum of the phonon spectral density, a series of wave packets is emitted yielding an even stronger degradation. The insight gained from our results will further help in designing an optimal preparation scheme for quantum dots.
\end{abstract}

\maketitle

\section{Introduction}
The interaction between acoustic phonons and electrons in self-assembled semiconductor quantum dots has been in the focus of research for quite some time now, unraveling several interesting phenomena: Prominent examples are the phonon side-bands induced by the acoustic phonons, which were first measured on a single quantum dot 20 years ago  \cite{besombes2001aco}. Around the same time, phonon emission from quantum dots was measured by bolometric techniques \cite{hawker1999energy,bellingham2001aco}. Another fascinating example is the reappearance regime which has been predicted for Rabi rotations in 2007 \cite{vagov2006hig,vagov2007non,vagov2012las} and measured for adiabatic rapid passage about 10 years later \cite{kaldewey2017dem}. The common conception that phonons are unwanted and should be avoided, because they lead to decoherence \cite{htoon2001car,alicki2004opt,krugel2005the,grodecka2007int,ramsay2010dam,kaer2013mic,gawarecki2013adi}, was being revisited since about 10 years ago, when phonon-assisted processes were proposed for the preparation of the exciton \cite{reiter2012pho,glassl2013pro} and subsequently experimentally implemented \cite{quilter2015pho,bounouar2015pho,ardelt2014dis,thomas2021bri}. More details on phenomena of the electron-phonon interaction in quantum dots can be found in recent reviews \cite{reiter2019dis,luker2019review}.

Nowadays, quantum dots are discussed as deterministic photon sources showing very promising results in the race towards the perfect single-photon source to be used in quantum technology \cite{santori2001tri,ding2016dem,schweickert2018ond, rodt2020det,arakawa2020pro,tomm2021bri,thomas2021rac}. To maximize the performance of the quantum dots and excitation protocols, any detrimental influences are under review \cite{naesby2008inf}. In particular, the electron-phonon interaction must be carefully studied to understand how to optimize quantum dots for photon generation. Phonons have a direct impact on the state preparation of quantum dot states \cite{luker2019review}, which directly affects the subsequent photon generation, but also the properties of the generated photons, e.g., the indistinguishability is influenced by the phonons \cite{close2012ove,kaer2013mic,thoma2016exp,carmele2019non,cosacchi2019emi,cosacchi2021acc}.

To obtain the optimal photon source, several preparation schemes have been developed in the past few years, which go beyond the usual Rabi rotations \cite{kamada2001exc,stievater2001rab,borri2002rab,forstner2003pho,machnikowski2004res, stufler2006two,ramsay2010dam}, including adiabatic rapid passage (ARP) \cite{wu2011pop,simon2011rob,debnath2012chi,debnath2013hig,wei2014det,kaldewey2017dem}, phonon-assisted preparation \cite{reiter2012pho,glassl2013pro,quilter2015pho,bounouar2015pho,ardelt2014dis}, pulse-optimized \cite{stefanatos2020rap} and two-color excitation schemes \cite{he2019coh,koong2021coh}. A radically new scheme, based on the swing-up of emitter population (SUPER) process has been proposed just this year \cite{bracht2021swi}. 

\begin{figure}
    \centering
    \includegraphics[width=0.48\textwidth]{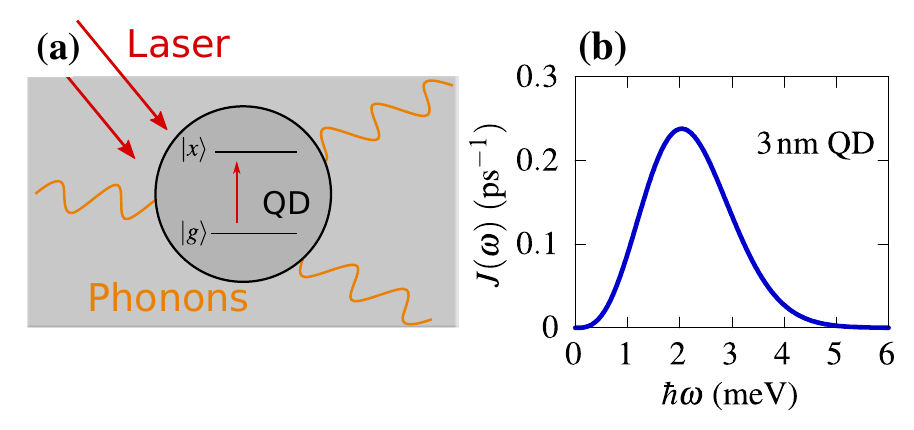}
    \caption{(a) Setup considered here. A spherical semiconductor quantum dot (QD) is excited by a short laser pulse and emits phonon wave-packets. (b) The phonon spectral density for a $\SI{3}{nm}$ spherical GaAs quantum dot.}
    \label{fig:schematic_plot}
\end{figure}

In this paper, we consider the generated phonons during state preparation. As schematically shown in \textbf{Figure~\ref{fig:schematic_plot}}(a), a quantum dot is excited by a laser pulse accompanied by the generation of phonons. We compare the newly proposed SUPER scheme with the resonant Rabi rotations and the ARP. All three schemes feature a coherent excitation process of the quantum dot. The SUPER scheme relies on a two-color excitation process, where both spectral components are several millielectronvolts below the ground state to exciton transition energy. Therefore, spectral filtering between the exciting pulses and the emitted photons can be easily implemented. 
The emission of phonon wave packets has so far been studied for Rabi oscillations but not for other preparation schemes like ARP or SUPER, which is done in the present paper.

\section{Theoretical background}
\subsection{Model and Hamiltonian}
For all schemes, we consider the excitation of a semiconductor quantum dot with circularly polarized laser pulses. In this case, the quantum dot can be described as a driven two-level system. The transition frequency between ground state $\ket{g}$ and excited state $\ket{x}$ is denoted by $\omega_0$ and the transition is driven by an external laser field $E(t)$ described in the usual dipole and rotating wave approximation. This results in a driving term $\Omega(t)=dE(t)/\hbar$ with $d$ being the dipole moment of the transition. The Hamiltonian including the carrier-light interaction reads
\begin{equation}
    H = \hbar\omega_0\ket{x}\bra{x} - \frac{\hbar}{2}\left(\Omega^*(t)\ket{g}\bra{x} + \Omega(t)\ket{x}\bra{g}\right).
\end{equation}
For the driving laser field, for each pulse $n$ a Gaussian shape is assumed with
\begin{equation}
  \Omega(t) = \Omega_n(t) e^{-i\phi_n(t)}, \quad\Omega_n(t) = \frac{\alpha_n}{\sqrt{2\pi\sigma_n^2}} e^{-(t-\tau_n)^2/(2\sigma_n^2)},\label{eq:electric_field_simple}
\end{equation}
where $\Omega_n(t)$ is the real pulse envelope centered around $\tau_n$, $\alpha_n=\int\Omega_n(t)\,dt$ is the pulse area and $\sigma_n$ the duration of the pulse, related to the full width at half maximum by $\text{FWHM}_n = \sqrt{2\ln 2}\sigma_n$. The expression for the phase $\phi_n(t)$ depends on the specific scheme that is considered. The important quantities for the dynamics are the occupation of the exciton state $f=\langle\ket{x}\bra{x} \rangle$ and the polarization $p=\langle\ket{g}\bra{x} \rangle$. For plotting the polarization, we change into a rotating frame with reference frequency $\omega_R$. 


\subsection{State preparation schemes}
In this paper, we will compare three different coherent excitation schemes, which have different conditions on the frequencies of the used laser pulses:\\
\begin{tabular}{ll}
    Rabi:&$\phi_{\text{Rabi}}(t) = \omega_0 t$\\
    ARP$^{\pm}$:&$\phi_{\text{ARP}}(t) = \omega_0 t \pm \frac{|a|}{2}t^2$\\
    SUPER:&$\phi_{\text{S1}}(t) = \omega_1 t,\quad \phi_{\text{S2}}(t) = \omega_2 t$
\end{tabular}\\
In the Rabi scheme, a resonant excitation with pulse area $\pi$ excites the system by stimulating half a Rabi oscillation. In the ARP scheme, the frequency of the laser pulse changes linearly described by the chirp parameter $a$, leading to an adiabatic evolution along the dressed states if the Landau-Zener criterion is fulfilled \cite{tannor2008int}. The SUPER scheme relies on the swing-up mechanism, and can be achieved by combining two pulses at the energies $\hbar\omega_1$ and $\hbar\omega_2$ \cite{bracht2021swi}, with both pulses being several millielectronvolts below the transition frequency. Note that such a pulse alone would not lead to an occupation change. The two pulses have different pulse areas $\alpha_j$ and pulse lengths $\sigma_j$ and can have a temporal separation $\tau$. For the SUPER scheme to be efficient, the energy difference between the two pulses should correspond to the Rabi frequency of the less-detuned pulse at its maximum \cite{bracht2021swi}
\begin{equation}
    |\omega_2 - \omega_1| =  \sqrt{\Omega_{1}^2(t=0) + (\omega_1-\omega_0)^2} = \Omega^{\text{R}}_{\text{diff}} \, . \label{eq:second_frequency} 
\end{equation}
By choosing a negative detuning, it is possible to excite within the band gap of the semiconductor structure, which can also be advantageous in experiment. A similar excitation mechanism can be achieved by modulation of the frequency or the amplitude of the exciting pulse \cite{shi2016pop,shi2021two,bracht2021swi}. We remark that this is different to the bichromatic excitation schemes \cite{he2019coh,koong2021coh}, where one frequency is detuned above the transition frequency and the other below. In the following we express the frequencies of the lasers in terms of the detuning compared to the transition frequency of the quantum dot $\Delta_{j} =\omega_j - \omega_0$. This translates to a detuning of zero for the Rabi scheme, a changing detuning $\Delta(t)=at$ for ARP and for the SUPER scheme both pulses are chosen to have $\Delta_j < 0$.

\subsection{Electron-phonon interaction}
In quantum dots, it is known that the preparation of the exciton is prone to disturbance by phonon effects \cite{luker2019review}. In particular, Rabi rotations and the ARP are damped in certain parameter regimes due to phonons. At low temperatures, the main source for the damping is the interaction of the excitonic system with longitudinal acoustic (LA) phonons, which has been confirmed in several joint experimental and theoretical studies \cite{ramsay2010dam,kaldewey2017dem}. The standard Hamiltonian for the phonons and the exciton-phonon interaction is given by
\begin{equation}
H_\text{ph}= \hbar \sum_{\vec{q}} \omega_{\vec{q}} b_{\vec{q}}^{\dagger} b_{\vec{q}}^{}+\hbar \ket{x}\bra{x} \sum_{\vec{q}} \left(g_{\vec{q}} b_{\vec{q}} + g_{\vec{q}}^* b_{\vec{q}}^{\dagger} \right)\,.
\end{equation}
The operator $b_{\vec{q}}^{\dagger} (b_{\vec{q}})$ is a bosonic operator that creates (annihilates) a phonon with wave vector $\vec{q}$ and energy $\hbar\omega_{\vec{q}}$ and is coupled to the exciton states via $g_{\vec{q}}$. For the LA phonons, we assume a linear dispersion relation. \\
A measure for the strength of the carrier-phonon interaction is given by the phonon spectral density defined as
\begin{equation}
J(\omega) = \sum_{\mathbf{q}} \left| g_{\mathbf{q}}\right|^2 \delta (\omega-\omega_{\mathbf{q}}) \, ,
\label{eq:spectral}
\end{equation}
Remembering that the coupling matrix element $g_ {\vec{q}}$ is composed of the bulk coupling matrix element and the form factor determined by the electronic wave functions, the phonon spectral density is non-monotonous as a function of energy. For the electronic properties it is sufficient to consider the phonon spectral density, while for the emission patterns of the phonons, the specific shape of the quantum dot \cite{luker2017pho}, as well as the exciton phonon coupling mechanism is relevant, e.g., the piezo-electric coupling leads to a highly anisotropic phonon emission even for spherical dots \cite{krummheuer2005pure}. To simplify our discussion, we concentrate on deformation potential coupling and a spherical quantum dot with a radius of $\SI{3}{\nano\meter}$ and standard GaAs parameters \cite{luker2017pho}. The resulting phonon spectral density is shown in Figure~\ref{fig:schematic_plot}(b). For these parameters the maximum is found at $\hbar\omega^{\text{ph}}_{\text{max}}=\SI{2.05}{meV}$. 
\\
The interaction with the phonons can theoretically be described on several levels including rate equations \cite{klassen2021opt}, polaron-transformed master equations \cite{mccutcheon2011gen,roy2011inf} or path-integral formulations \cite{vagov2011rea} and generating functions \cite{axt2003ana}. Here, we use a correlation expansion approach to simulate the coupled quantum dot-phonon system \cite{ahn2005res,krugel2005the}. In the correlation expansion, the infinite hierarchy of equations of motions, which builds up due to the many-body nature of the electron-phonon interaction, is truncated by neglecting correlations starting from a certain order. It has been shown that omitting correlations of three particles, corresponding to a fourth order Born approximation, gives an adequate description for excitation schemes of quantum dots \cite{krugel2006back}. We stress within our method, that non-Markovian effects of the dynamics are adequately described and the polaron formation can be well monitored. The correlation expansion has been shown to compare well with the numerically exact path integral formalism \cite{vagov2011dyn}. An advantage of the correlation expansion is that the phonon dynamics is explicitly taken into account, such that also phonon properties like fluctuation dynamics can directly be extracted from the calculations \cite{reiter2011flu,wigger2014ene}.\\
To evaluate the influence of phonons on the preparation schemes, we also analyze the corresponding phonon dynamics. It is well established that the ultra-fast excitation of a quantum dot results in the formation of the polaron, visible as a stationary lattice distortion in the vicinity of the quantum dot, accompanied by the emission of a phonon wave packet \cite{krummheuer2005cou,jakubczyk2016imp,vanacore2017ult}. A similar dynamics also holds for finite pulses, where the phonon dynamics depends on the excitation conditions. The dynamics of the phonons can be described by the lattice displacement ${\mathbf{u}}({\mathbf{r}}) = \langle \hat{{\mathbf{u}}}({\mathbf{r}})\rangle $ defined as the expectation value of the operator \cite{wigger2014ene}
\begin{equation}
\hat{ {\mathbf{u}} }({\mathbf{r}}) = - i \sum_{\mathbf{q}} \sqrt{\frac{\hbar}{2\rho V
\omega_{\mathbf{q}}}}\left(\hat{ b}_{\mathbf{q}} e^{i{\mathbf{q}}\cdot{\mathbf{r}}} -
\hat{ b}_{\mathbf{q}}^\dag e^{-i{\mathbf{q}}\cdot{\mathbf{r}}}\right)\frac{{\mathbf{q}}}{q} .
\label{eq:u}
\end{equation}
In the case of a spherically symmetric quantum dot, also the lattice displacement is radial, i.e.,  $\langle \hat{{\mathbf{u}}}({\mathbf{r}})\rangle = u(r)
{\mathbf{e}}_r$, and we will consider the scaled quantity
\begin{equation}
\tilde{u}(r)=\left( \frac{r}{1\,{\rm{nm}}} \right) u(r) \,,
\label{eq:utilde}
\end{equation}
which compensates for the radial attenuation of the lattice displacement.\\
We will perform all calculations at zero temperature $T=\SI{0}{K}$ to focus on the emission of the phonon wave packets.
\section{Results}
\subsection{Phonon damping during state preparation}
\begin{figure*}
    \centering
    \includegraphics[width=\textwidth]{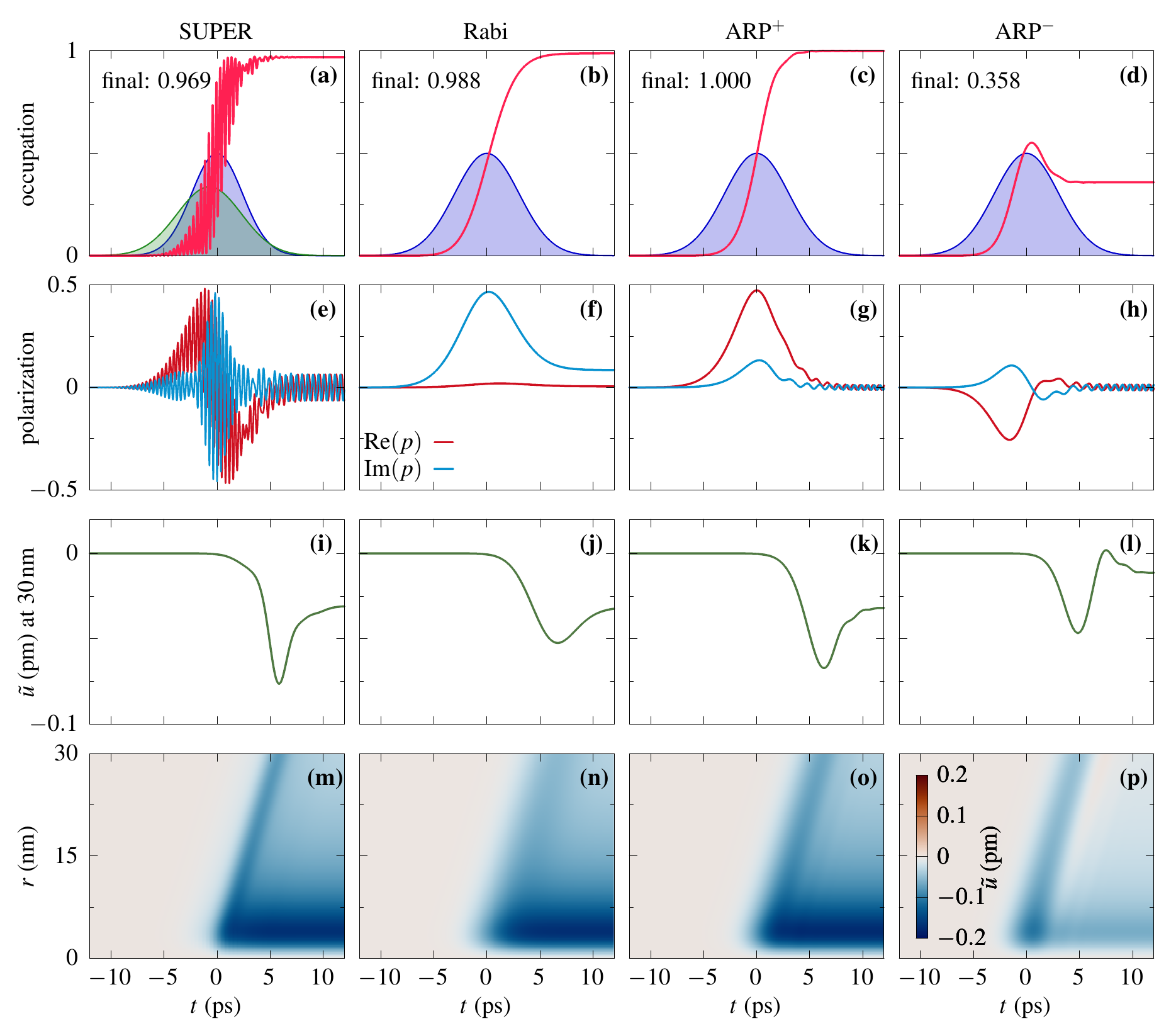}
    \caption{Phonon impact on (a) SUPER, (b) Rabi, (c) ARP$^+$ and (d) ARP$^-$ scheme: Occupation of the excited state (red line) and the corresponding pulses (blue and green shapes). (e,f,g,h) The real and imaginary part of the polarization as a function of time. (i,j,k,l) The scaled lattice displacement $\tilde{u}$ at $r=\SI{30}{nm}$ and (m,n,o,p) $\tilde{u}(r)$ in a color plot as a function of distance $r$ from the quantum dot and time $t$.}
    \label{fig:super_compare}
\end{figure*}

Without phonons, all three schemes lead to a perfect inversion of the two-level system, i.e., after the laser pulses the final occupation of the exciton state is unity. The dynamics of the exciton occupation as a function of time, under the influence of phonons, is plotted in \textbf{Figure~\ref{fig:super_compare}} for (a) the SUPER scheme, (b) the Rabi scheme, and (c,d) the ARP$^\pm$ scheme with positive and negative chirp respectively. The shaded areas indicate the pulse shapes with the blue pulses normalized to $0.5$. In the second row the corresponding polarizations are plotted. Note that the polarizations are shown in a rotating frame with $\omega_R=\omega_0$ for the Rabi scheme, $\omega_R(t)=\omega_0\pm at $ for the ARP scheme and $\omega_R=\omega_1$ in the SUPER scheme.

Let us start with the most established scheme, i.e., the Rabi scheme. In this case, the pulse parameters are $(\sigma=\SI{3.04}{ps},\alpha=\pi)$. The occupation [cf. Figure~\ref{fig:super_compare}(b)] shows a smooth increase of the occupation during the pulse, while the imaginary part of the polarization increases and decreases. Due to the phonons, the final occupation of $1$ is not achieved, but only $0.988$, accompanied by a small Re($p$) and Im($p$). For the Rabi scheme, it is well studied that the final occupation depends on the exciting pulse parameters \cite{wigger2014ene}. With a pulse duration of $\SI{3.04}{ps}$, we are close to experimentally used parameters \cite{ding2016dem,gustin2018pul}.

For the ARP$^+$ scheme [cf. Figure~\ref{fig:super_compare}(c)] we use the following pulse parameters: ($\sigma=\SI{3.04}{ps},\alpha=5\pi, a=\SI{1}{ps^{-2}}$). The occupation dynamics also increases smoothly up to one, while in the polarization we find that Re($p$) becomes much stronger than Im($p$). The impact of the phonons on the final occupation is so small, that the occupation reaches $1$ within the given accuracy. We remind the reader that the phonon influence strongly depends on the sign of the chirp \cite{luker2012inf}. Accordingly, we also look at the dynamics of ARP$^{-}$, i.e., the chirp parameter is now $a=\SI{-1}{ps^{-2}}$. Here, phonons have a great impact on the final occupation [cf. Figure~\ref{fig:super_compare}(d)], which only reaches a values of $0.358$ after the pulse. The asymmetry of the phonons regarding the sign of the chirp has been confirmed several times in experiments \cite{debnath2012chi,debnath2013hig,mathew2014sub,kaldewey2017coh,kaldewey2017dem,ramachandran2020sup}.

For both the Rabi and ARP scheme, the phonon influence on the state preparation can be estimated via the dressed states. The dressed states are the eigenstates of the coupled dot-light system with the energies $E_{\pm}$. When the energy splitting of the dressed states is such that it overlaps with the phonon spectral density, transitions between the dressed states can take place. At low temperatures, only phonon emission is possible, which is the reason for the sign-asymmetry in the ARP scheme. Accordingly, if the splitting of the dressed states does not lie within the phonon spectral density, the reappearance regime is entered, where the phonons are not efficiently coupled anymore \cite{vagov2007non,kaldewey2017dem}. For the quantum dot parameters chosen here, this would be beyond $\SI{5}{meV}$.

With this in mind, we now turn to the SUPER scheme. We use two pulses with $\hbar\Delta_1 = \SI{-8}{meV}$, $\sigma_1=\SI{2.4}{ps}$, $\alpha_1=22.65\pi$ and $\hbar\Delta_2 = \SI{-19.163}{meV}$, $\sigma_2=\SI{3.04}{ps}$, $\alpha_2=19.29\pi$ and a time delay of $\tau=\SI{-0.73}{ps}$ between the pulses, such that $\hbar\Omega^{\text{R}}_{\text{diff}}=\SI{11.163}{meV}$. These pulse parameters are the same ones as used in Ref.~\cite{bracht2021swi}. The SUPER scheme relies on the idea, that the occupation is gradually increased by combining different Rabi frequencies in a clever way. Accordingly, we see fast oscillations with $\Omega^{\text{R}}_{\text{diff}}$ in the occupation [cf. Figure~\ref{fig:super_compare}(a)], while the mean value increases from zero to almost one. It is also interesting to look at the polarization, which also displays the fast oscillation with $\Omega^{\text{R}}_{\text{diff}}$ for the considered rotating frame. During these oscillations, Re$(p)$ is pronounced at the beginning and end of the pulse, while Im$(p)$ is strong at the pulse maximum. \\
The final occupation under the influence of phonons then reaches a value of $0.969$. This is somewhat surprising, because the detuning of both pulses with $\SI{-8}{meV}$ and $\SI{-19.163}{meV}$ is beyond the phonon spectral density. On the other hand, the picture of the dressed states cannot be easily applied to the SUPER scheme. Due to the strong modifications of the dressed states resulting from both pulses, one cannot assume that an adiabatic evolution along the dressed states takes place, but rather diabatic transitions between the dressed states occur. Nevertheless, further parameter studies could even lead to higher final occupations using the SUPER scheme.

\subsection{Wave packet emission}
To get more insight into the phonon influence on the different preparation schemes, we look at the lattice displacement induced by the different schemes. For this, we plot the scaled lattice displacement $\tilde{u}$ as function of distance $r$ from the quantum dot and time $t$ in Figure~\ref{fig:super_compare}(m,n,o,p). Additionally, we plot a cut at $r=\SI{30}{nm}$ in panels (i,j,k,l). In all cases, we see the formation of the polaron as a strong negative displacement in the vicinity of the quantum dot. Note that the confinement length of the quantum dot of $\SI{3}{nm}$ corresponds to the width of the electronic wave function. We find that the polaron depends mainly on the final occupation and accordingly, so the displacement patterns of the polaron for the SUPER, Rabi and ARP$^{+}$ schemes are almost the same. Only for the ARP$^{-}$ scheme an occupation of only $0.358$ is achieved, hence, the polaron is less pronounced. We will not consider the ARP$^{-}$ anymore in the following, because it does not lead to a high final occupation of the excited state.

\begin{figure}
    \centering
    \includegraphics{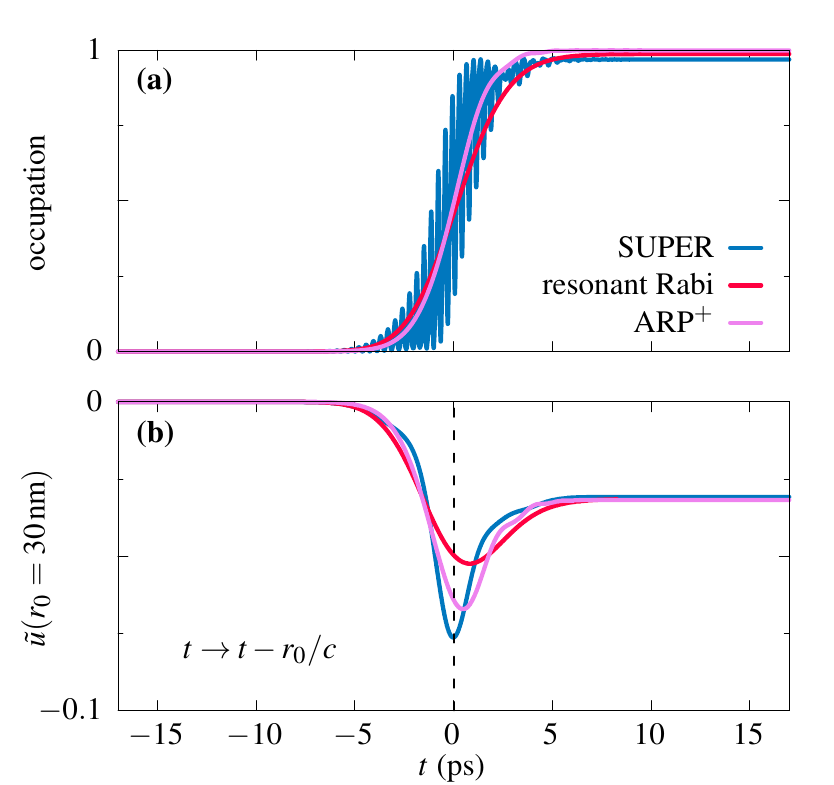}
    \caption{(a) Occupation of the exciton state for the three different schemes. (b) Scaled lattice displacement $\tilde{u}$ at $\SI{30}{nm}$ for the three schemes, shifted by $r_0/c_s$, where $c_s$ is the sound velocity.}
    \label{fig:cut_30nm}
\end{figure}

Additionally, during the excitation a phonon wave packet is emitted into the surrounding area of the quantum dot, visible as a diagonal line in the color plots. Interestingly, the form of the wave packet is  rather different for the different excitation schemes, which can be well seen in Figure~\ref{fig:super_compare}(i,j,k,l), showing cuts of the lattice displacement at $r=\SI{30}{nm}$. To compare the wave packet form for the different excitation schemes in detail, we contrast the scaled lattice displacement $\tilde{u}$ at $r_0=\SI{30}{nm}$ for the three schemes in \textbf{Figure~\ref{fig:cut_30nm}}(b), while the respective occupation dynamics are shown again in (a). The wave packets are seen around $t-r_0/c_s=\SI{0}{ps}$, i.e., shifted by the distance divided by the velocity of sound $c_s$ in the material. For longer times the lattice displacement reaches a plateau value which stems from the flank of the polaron and is therefore similar for all schemes.\\
Comparing the three cases, the minimum of $\tilde{u}$, corresponding to the largest lattice displacement, occurs at different times for each scheme. The difference of $\SI{-0.73}{ps}$ between the Rabi and SUPER scheme can be traced back to the pulse with $\omega_2$ arriving by $\tau=\SI{-0.73}{ps}$ earlier. Nonetheless the wave packet of the ARP$^+$ has its maximum earlier than the Rabi scheme, though the temporal maximum of both pulses occur at the same time. \\
The resulting wave packets also differ in their width, though the occupation dynamics show a rather similar rise-time. The resonant Rabi scheme features the broadest wave packet with the smallest amplitude, while in the SUPER scheme a rather narrow wave packet is emitted.

\subsection{SUPER at efficient phonon coupling regime}

\begin{figure}
    \centering
    \includegraphics[width=0.48\textwidth]{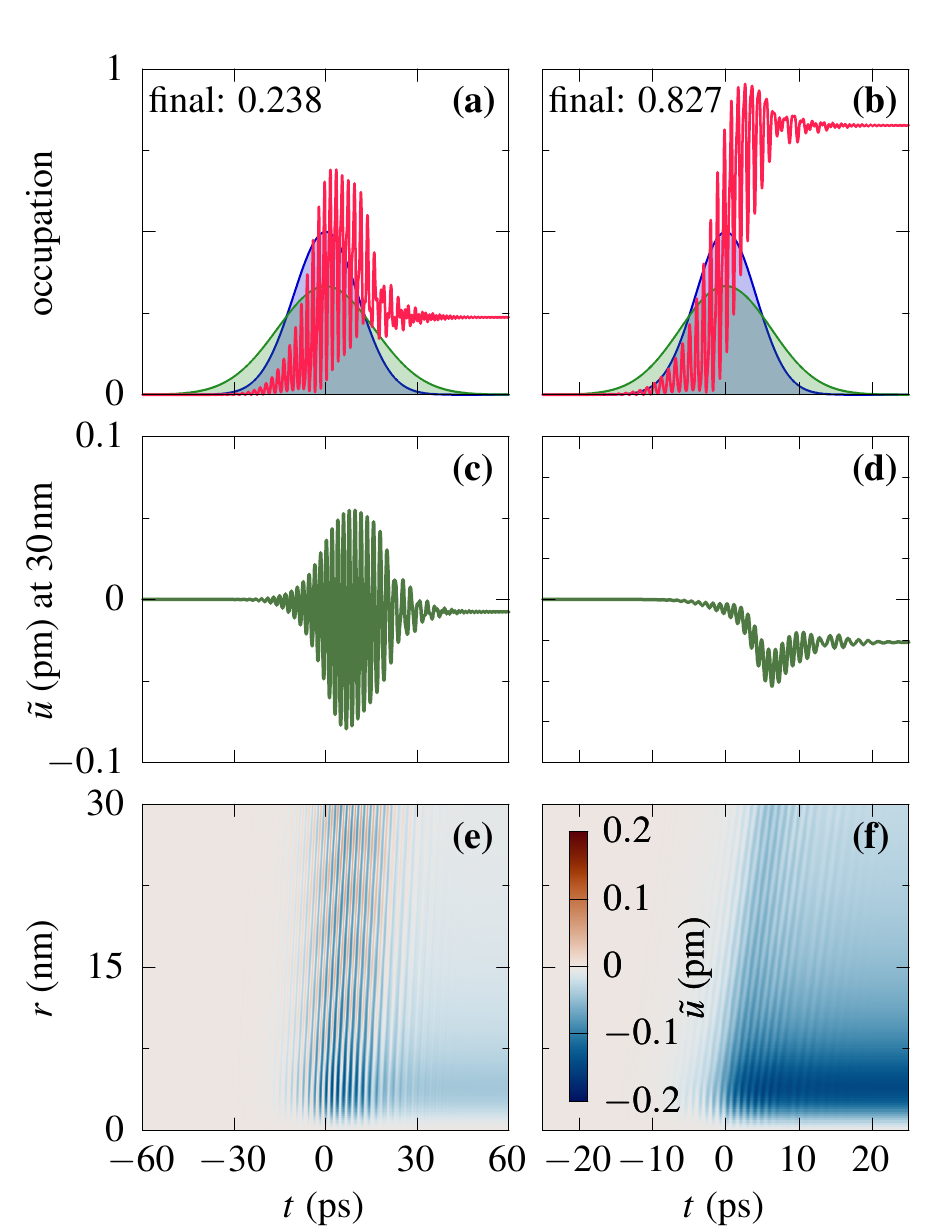}
    \caption{Phonon impact on the exciton occupation for the SUPER scheme using (a) $\hbar \Delta_1=\SI{-1.500}{meV},  \hbar\Delta_2=\SI{-3.614}{meV}$ (b) $ \hbar \Delta_1=\SI{-3.000}{meV},  \hbar\Delta_2=\SI{-7.231}{meV}$ (e,f) the lattice displacement and (c,d) the corresponding cut at $\SI{30}{nm}$}
    \label{fig:super_phonons}
\end{figure}

Keeping in mind that the phonon spectral density has its maximum at just a few millielectronvolts, the question arises, if the fast oscillations of the occupation in the SUPER scheme can couple to the phonons efficiently. To study this, we consider much smaller detunings which correspond to frequencies in the range of the maximum of the spectral density, i.e., in the range of about $1-3$ meV. Without phonons, also in this parameter regime, a complete population inversion can be achieved due to the SUPER mechanism \cite{bracht2021swi}. Taking into account phonons, we show two cases of the SUPER scheme in \textbf{Figure~\ref{fig:super_phonons}} with (a) $\hbar\Delta_1=\SI{-1.500}{meV},\hbar\Delta_2=\SI{-3.614}{meV}$ (other pulse parameters $\sigma_1=\SI{11.75}{ps}, \sigma_2=\SI{17.63}{ps}, \alpha_{1,2}=21.23\pi$ and $\tau=0$) and (b) $\hbar\Delta_1=\SI{-3.0}{meV},\hbar\Delta_2=\SI{-7.231}{meV}$  (other pulse parameters $\sigma_1=\SI{5.87}{ps}, \sigma_2=\SI{8.80}{ps}, \alpha_{1,2}=21.23\pi$ and $\tau=0$).

In the first example shown in Figure~\ref{fig:super_phonons}(a) the energetic difference $\hbar\Omega^{\text{R}}_{\text{diff}} = \SI{2.114}{meV}$ is close to the maximum of the phonon spectral density at $\hbar\omega^{\text{ph}}_{\text{max}}=\SI{2.05}{meV}$. In the dynamics of the occupation, we see that at the beginning of the pulses the SUPER scheme sets in, leading to a rise of the mean occupation. Then, however, the occupation does not reach one, but goes to a stationary value of $0.238$. Also the phonon dynamics looks rather different compared to the previous case shown in Figure~\ref{fig:super_compare}(m). In Figure~\ref{fig:super_phonons}(e) it can be seen that a series of small wave packets is emitted during the action of the pulses. The sign of the wave packets is alternating between positive and negative, as can best be seen in the corresponding cut at $r=\SI{30}{nm}$ in Figure~\ref{fig:super_phonons}(c). The sequence of emitted wave packets corresponds to the fast oscillations seen in the occupation, i.e., its frequency is  $\Omega^{\text{R}}_{\text{diff}}$. This is in agreement with findings on the wave packet emission from constantly and resonantly driven quantum dots, where each Rabi cycle leads to an emission of a wave packet \cite{wigger2014ene}.

In the second example shown in Figure~\ref{fig:super_phonons}(b) we take higher detunings, such that only the frequency of one pulse still is within the range of the phonon spectral density, while the other is already beyond it. The difference between the pulse frequencies is $ \hbar \Omega^{\text{R}}_{\text{diff}} = \SI{4.231}{meV}$, which is at the upper edge of the phonon spectral density. In the occupation we find that the phonons degrade the preparation fidelity to $0.827$. When looking at the wave packet dynamics displayed in Figure~\ref{fig:super_phonons}(f), we find that a wave packet is formed with an additional fast oscillation of a small amplitude, which becomes particularly clear at the cut shown in Figure~\ref{fig:super_phonons}(d). The emission frequency of the phonon wave packets again corresponds to the fast oscillation of the occupation with $\Omega^{\text{R}}_{\text{diff}}$. This finding is consistent 
with considerations regarding the reappearance regime for Rabi rotations or ARP \cite{vagov2007non,kaldewey2017dem}, that beyond the phonon spectral density an efficient phonon coupling is suppressed. Nonetheless, even at high detunings as shown in Figure~\ref{fig:super_compare}(m), a phonon wave packet is emitted and a slight degradation of the occupation takes place. 

\section{Conclusion}
In conclusion, we have discussed the phonon impact on different excitation schemes for quantum dots. Depending on the excitation scheme, the phonons lead to a more or less pronounced degradation of the final occupation, while phonon wave packets are always emitted during the excitation process. We have put special emphasis on the recently proposed SUPER scheme, which uses an excitation with two negatively detuned pulses to excite the quantum dot. Relying on the swing-up process, the occupation dynamics of the SUPER scheme during the pulses feature both a fast oscillation and a steady rise of the occupation. If the fast oscillation is such that it resonates with the phonons, a series of wave packets is emitted with this distinct frequency. If the frequency of the fast oscillation is too high, only a single, but broader wave packet is emitted. Our results give insight to how the scheme can be used to result in minimal phonon influence, by choosing the difference of the detunings to be outside the phonon spectral density. 
\acknowledgements
T.K.B. and D. E. R. thank the German Research Foundation DFG for financial support through the project No. 428026575.
\bibliography{main}
\end{document}